\documentclass[cits]{PoS}
\title{Extracting Hadron Parameters from Dispersive Sum
Rules}\ShortTitle{Extracting Hadron Parameters from Dispersive Sum
Rules}\author{\speaker{Wolfgang Lucha}\\Institute for High Energy
Physics, Austrian Academy of Sciences, Nikolsdorfergasse 18,
A-1050 Vienna, Austria\\E-mail: \email{Wolfgang.Lucha@oeaw.ac.at}}
\author{Dmitri Melikhov\\Institute for High Energy Physics,
Austrian Academy of Sciences, Nikolsdorfergasse 18, A-1050 Vienna,
Austria, and\\D.~V.~Skobeltsyn Institute of Nuclear Physics,
Moscow State University, 119991, Moscow, Russia\\E-mail:
\email{dmitri\_melikhov@gmx.de}}
\author{Silvano Simula\\INFN, Sezione di Roma III, Via della Vasca
Navale 84, I-00146, Roma, Italy\\E-mail:
\email{simula@roma3.infn.it}}

\abstract{Puzzled or surprised by the almost incredible accuracy
occasionally claimed in the literature to be achievable for
numerical outcomes of QCD sum-rule analyses, we scrutinized the
usual procedure employed for the extraction of the parameters of
individual bound states from dispersive sum~rules by taking
advantage of the exact solvability of a quantum-mechanical
harmonic-oscillator model: It turns out that the
\emph{determination of the ground-state parameters} (that is,
decay constant and~form factor) \emph{by requiring independence
from the Borel mass} in its stability window \emph{does not
necessarily yield their exact numerical values} \cite{LMS:PRD76,
LMS:QCD@Work07,LMS:YaF,LMS:PLB657,LMS:Hadron07,LMS:combine}. For
instance, the comparison of the sum-rule predictions for
bound-state parameters with their numerical values known precisely
in our harmonic-oscillator model reveals that standard sum-rule
procedures underestimate the ground-state decay constant by some
4\% and its form factor by almost 15\%; such systematic
uncertainties cannot be inferred from our correlators' accuracy
better than 1\% in the window of Borel stability: they are
uncontrollable.}

\FullConference{8th Conference Quark Confinement and the Hadron
Spectrum\\September 1--6, 2008\\Mainz, Germany}

\begin{document}The idea behind this sequence of studies at
quantum-physics level \cite{LMS:PRD76,LMS:QCD@Work07,LMS:YaF,
LMS:PLB657,LMS:Hadron07,LMS:combine} is simple and elegant: For a
harmonic-oscillator model defined by the nonrelativistic
Hamiltonian $H=\mathbf{p}^2/2m+m\omega^2\mathbf{x}^2/2$
correlators, such as the polarization $\Pi(E)\equiv
\langle\mathbf{x}_{\rm f}=\mathbf{0}|(H-E)^{-1}|\mathbf{x}_{\rm i}
=\mathbf{0}\rangle,$ are known at both ``hadron'' and ``quark''
levels, where the counterpart of the operator product expansion in
QCD is easily found; at ``hadron'' level, the Borel transform of
$\Pi(E),$ with Borel mass $\mu,$ is $\Pi(\mu)=\sum_{n=0}^\infty
R_n\exp(-E_n/\mu)$ with (exact) energies $E_0=3\omega/2,$
$E_1=7\omega/2,$ \dots\ and decay constants
$R_n\equiv|\Psi_n(\mathbf{x}=\mathbf{0})|^2$. To extract from the
quark level the ground-state parameters $E_0$ and $R_0$
numerically, \emph{duality}, i.e., equality of the continuum
contributions above a threshold energy is assumed. Equating the
remainders of ``hadron'' and ``quark'' expressions yields a
\emph{sum rule}. This is best illustrated graphically (Fig.~1).
However, the threshold's $\mu$-dependence induces uncontrolled
uncertainties into this approach; without specifying this
dependence sum-rule predictions are not reliable but plagued by
considerable systematic errors.

\begin{figure}[h]\begin{center}
\includegraphics[width=6.9cm]{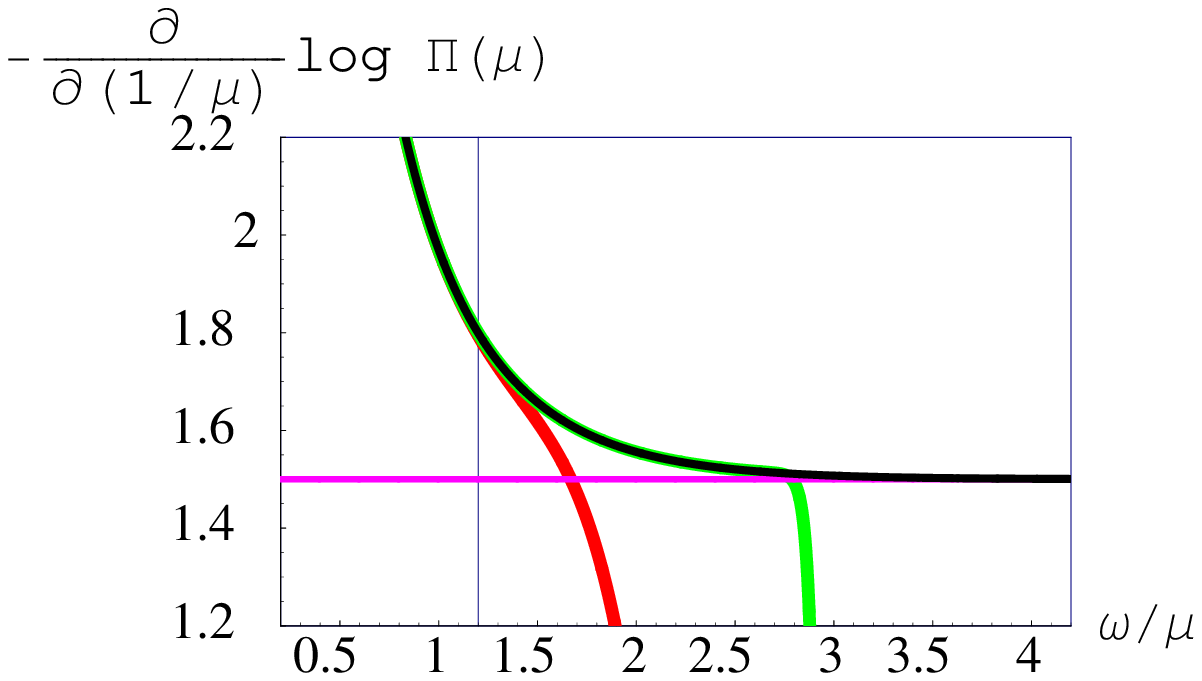}
\includegraphics[width=6.9cm]{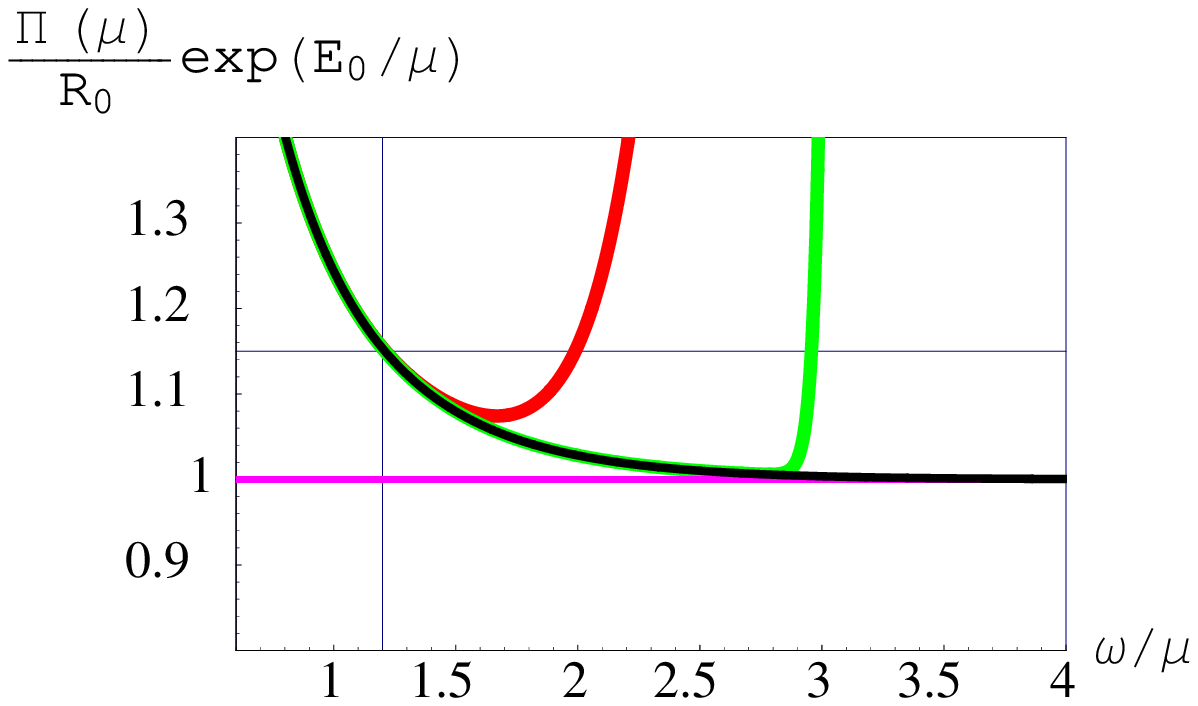}\end{center}\vspace{-3ex}
\caption{Extraction of the ground-state energy $E_0$ and decay
constant $R_0$ of our model by either considering the exact
expression (black) for $\Pi(\mu),$ or retaining 4 (red) and 100
(green) terms in its expansion for large~$\mu.$}\end{figure}

\noindent{\bf Acknowledgements.} D.~M.\ gratefully acknowledges
financial support from the Austrian Science Fund (FWF) under
projects P17692 and P20573, and from the RFBR under project
07-02-00551.

\end{document}